\documentclass{elsart4-1}

  \usepackage{graphicx}

\usepackage{amssymb}

\usepackage[english,francais]{babel}

\newtheorem{e-proposition}[theorem]{Proposition}

\newtheorem{e-definition}[theorem]{Definition\rm}

\renewcommand{\theequation}{\arabic{equation}}
\def\og{\leavevmode\raise.3ex\hbox{$\scriptscriptstyle\langle\!\langle$~}}
\def\fg{\leavevmode\raise.3ex\hbox{~$\!\scriptscriptstyle\,\rangle\!\rangle$}}

\newcommand{\cF}{\ensuremath{\mathcal F}}

\newcommand{\bbA}{{\ensuremath{\mathbb A}} }

\newcommand{\bbF}{{\ensuremath{\mathbb F}} }

\newcommand{\gb}{\beta}

\newcommand{\gep}{\varepsilon}       
\newcommand{\gp}{\varphi}
\newcommand{\gr}{\rho}

\newcommand{\go}{\omega}

\newcommand{\gl}{\lambda}

\newcommand{\gs}{\sigma}

\newcommand{\bra}{\langle}
\newcommand{\ket}{\rangle}

\newcommand{\ggr}{r}
\newcommand{\eps}{\varepsilon}

\begin{document}

\centerline{Physics or Astrophysics/Header}
\begin{frontmatter}


\selectlanguage{english}
\title{
Cumulants and large deviations of the current through non-equilibrium steady states
}


\selectlanguage{english}
\author[authorlabel1]{T. Bodineau},
\ead{bodineau@math.jussieu.fr}
\author[authorlabel2]{B. Derrida}
\ead{derrida@lps.ens.fr}

\address[authorlabel1]{
Universit{\'e}s Paris VI $\&$ VII,
Laboratoire de Probabilit{\'e}s et Mod{\`e}les
Al{\'e}atoires, CNRS-UMR 7599, 
4 place Jussieu, Case 188, F-75252 Paris Cedex 05, France}
\address[authorlabel2]{Laboratoire de Physique Statistique, Ecole Normale
Sup{\'e}rieure,
24 rue Lhomond, 75231 Paris Cedex 05, France}

\thanks{We thank H. Spohn for useful suggestions. We acknowledge the support of the ACI-NIM 168 {\it Transport Hors Equilibre} of the Minist\`ere de l'Education Nationale, France.}


\medskip
\begin{center}
{\small Received *****; accepted after revision +++++}
\end{center}

\begin{abstract}
Using a generalisation of the detailed balance for systems maintained out of equilibrium by contact with 2 reservoirs at unequal temperatures or at unequal densities, we recover the fluctuation theorem for the large
deviation funtion of the current.
For large diffusive systems, we show how the large deviation funtion of the current can be computed using a simple additivity principle.
The validity of this additivity principle and the occurence of phase transitions are discussed in the framework of the macroscopic fluctuation theory.


\vskip 0.5\baselineskip

\selectlanguage{francais}
\noindent{\bf R\'esum\'e}
\vskip 0.5\baselineskip
\noindent
{\bf Cumulants et grandes d\'eviations du courant dans des \'etats stationnaires hors \'equilibre.}

En g\'en\'eralisant la relation de bilan d\'etaill\'e \`a des syst\`emes maintenus  hors \'equilibre par 
contact avec deux r\'eservoirs \`a des  temp\'eratures  ou \`a des densit\'es diff\'erentes, nous retrouvons le th\'eor\`eme
de fluctuations pour la fonction de grandes d\'eviations du courant.
Pour de grands syst\`emes diffusifs, nous montrons comment la fonction de grandes d\'eviations du courant 
peut \^etre calcul\'ee simplement \`a l'aide d'un principe d'additivit\'e.
La validit\'e de ce principe d'additivit\'e et l'existence de transitions de phase sont discut\'ees
dans le cadre d'une th\'eorie des fluctuations \`a l'\'echelle macroscopique.


 \vskip 0.5\baselineskip
\noindent{\small{\it Mots-cl\'es~:} Syst\`emes hors \'equilibre~; Fluctuations du courant~;
Bilan d\'etaill\'e g\'en\'eralis\'e}}
\end{abstract}
\end{frontmatter}


\selectlanguage{english}
\section{Introduction}
\label{}
A physical system in contact with two heat baths at unequal temperatures
$T_a$ and $T_b$
is one of the simplest situations  for which one can  observe a
non-equilibrium steady state.
\begin{figure}[ht]
\centerline{{\includegraphics[width=7cm]{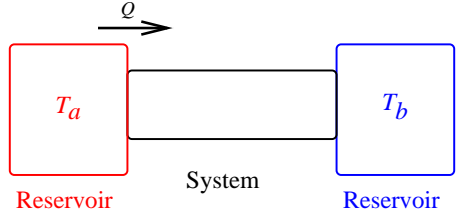}}}
\caption{ A system maintained in contact with two heat baths at unequal
temperatures reaches in the long time limit  a non-equilibrium steady state}
\label{fig:model}
\end{figure}
At equilibrium, i.e. when the two heat baths are at the same temperature
($T_a=T_b=T$),
the probability $P( C)$ of finding the system in a  given
microscopic configuration $C$ is given by the usual Boltzmann-Gibbs weight
\begin{equation}
P( C) = Z^{-1} \exp \left[ - {E( C) \over k T} \right]
\label{Boltzmann}
\end{equation}
where $E(C)$ is the internal energy of the system in configuration $C$.
Over the whole 20th century,
  studies in equilibrium statistical
mechanics have been based on this  expression or its microcanonical
counterpart, and the great success of
the theory was to show that  (\ref{Boltzmann}) was the right starting
point to explain the equilibrium properties of a large variety of
physical  systems 
(fluids, magnets, alloys, plasmas,....) and to understand all kinds of
effects,  in particular phase transitions and critical phenomena.
A very simplifying aspect of (\ref{Boltzmann}) is  that it  depends neither  on
the precise nature of the
coupling with the heat bath (at least when this coupling is weak) nor on the detailed dynamics of the system.

As soon  as the two temperatures $T_a$ and $T_b$ are different \cite{LLP}, there is not such a simple 
expression \cite{ruelle1,ruelle2} which generalizes (\ref{Boltzmann}) for the
steady state weights $P( C)$ of the microscopic configurations.  In fact for a non-equilibrium system, the steady state measure $P(C)$
depends in general on the precise description of the dynamics of
the system, of the heat baths and on their couplings. 
So far the exact expression of these weights is  known only for a few non-equilibrium models
\cite{DF,KLS,DEHP,SD}.

\medskip

In addition to the steady state weights,  one might be interested in 
the flow of energy through the system. For an interval of time $t$, one
may  consider the energy $Q_t$, the energy transfered from the heat bath
at temperature $T_a$ to the system.  In the steady
state, this energy fluctuates  and one might try to predict its various
{\it cumulants} $\langle Q_t^n \rangle_c$ or {\it its large deviation
function} $\cF(j)$ defined as
\begin{equation} 
{\rm Pro} \left ( {Q_t \over t}= j \right ) \sim
\exp [ -  t \cF(j) ] \ \ \ \ \ \ \ \ \ {\rm for \  large  } \ t
\label{F(j)}
\end{equation}
We refer to \cite{DZ,E} for a full account on the large deviation theory.
Note also that other definitions of the current distribution have been considered in
\cite{DSt1,DSt2}.

\begin{figure}[b]
\centerline{{\includegraphics[width=7cm]{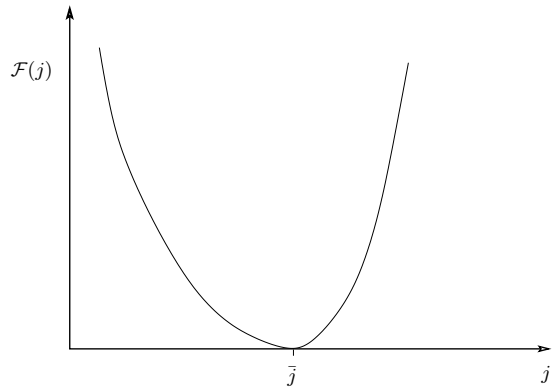}}}
\caption{A typical shape of the large deviation function $\cF(j)$ which vanishes at the typical current 
$\bar j$.}
\label{fig:fj}
\end{figure}

The whole distribution of $Q_t$ and a fortiori its cumulants depend in principle on the initial configuration $C_{\rm initial}$, on the final configuration $C_{\rm final}$ and on the place where the flux of energy is measured.
However
if the internal energy of the system is bounded ($\max_C |E(C)| < \infty$), the  cumulants of $Q_t$ (in the long time limit) and the large  deviation function $\cF(j)$  do not
depend on where the flow of energy is measured.
In particular, if one measures the flux of energy between the system and the other heat bath,
the large deviation function $\cF(j)$  is unchanged.
Also if the system relaxes faster than the time $t$ over which $Q_t$ is measured, the cumulants   
$\langle Q_t^n \rangle_c$ divided by $t$ and the large deviation function $\cF(j)$ do not depend on the initial and final configurations $C_{\rm initial},C_{\rm final}$.
In fact in this case it is elementary to verify that $\cF(j)$ is {\it convex},
that is if  $0 \leq \alpha \leq 1$
\begin{equation}
\cF \big( \alpha j_1 + (1- \alpha) j_2 \big) 
\leq 
 \alpha \cF(j_1) + (1- \alpha) \cF(j_2) 
\label{convexity} 
\end{equation}
as the probability distribution $  {\rm Pro}( Q_t|C_{\rm initial},C_{\rm final}) $ of $Q_t$,
given the initial and final configurations $C_{\rm initial}$ and $C_{\rm final}$, satisfies
\[  {\rm Pro}(  Q |C_{\rm initial},C_{\rm final})  = \sum_{C_{\tau}} \sum_q  {\rm Pro}(
 q |C_{\rm initial},C_{\tau})   {\rm Pro}( Q-q|C_{ \tau
},C_t)  \geq  {\rm Pro}(
  q |C_{\rm initial},C_{\tau})   {\rm Pro}( Q-q|C_{ \tau
},C_{\rm final}) \] 
which leads to (\ref{convexity})  in the long time limit when $\tau = \alpha t$,
$q= j_1 \alpha t$, $Q-q = j_2 (1- \alpha) t$.
The importance of convexity was understood in \cite{bdgjl} (see Section \ref{sec: macroscopic fluctuation} below).

It is sometimes easier to work with the generating function of $Q_t$. For large $t$ one has
\begin{equation}
 \left \langle e^{ \lambda Q_t} \right\rangle \sim e^{t \mu(\lambda)} 
\label{generating-function}
\end{equation}
where $\bra \cdot \ket$ denotes the expectation over the dynamics 
and $\mu(\lambda)$ is the Legendre transform of the large deviation function $\cF$
\begin{equation}
\mu(\lambda) = \max_j [j \lambda - \cF(j)] 
\label{mu(lambda)} 
\end{equation}
From the knowledge of $\mu(\lambda)$, one can often determine the cumulants of $Q_t$ in the
long time limit  by
\begin{equation}
\lim_{t \to \infty} { \langle Q_t^n \rangle_c  \over t} = \left. {d^n \mu(\lambda) \over d \lambda^n} \right|_{\lambda=0} 
\label{cum} 
\end{equation}
This relation is based on the assumption that the order of the limits $t \to \infty$ and $\lambda \to 0$  can be exchanged. One can show that these limits can be exchanged only for very few examples, although one believes that the assumption remains valid for general diffusive systems. There are however cases where these limits do not commute and for which the moments of the fluctuations cannot be deduced from the knowledge of the large deviation function  \cite{dls3}.

\section{Generalized detailed balance and the fluctuation theorem}

In principle determining the evolution of $Q_t$ requires the integration
of the evolution  equations of the system in presence of the heat baths.
This is  a difficult task, in particular because the heat baths are often
described by an infinite number of degrees of freedom.
Nevertheless, it can be shown in some cases that integrating the variables 
of the heat baths leads to effective reservoirs with stochastic noise.
We refer to \cite{EPR,BLR} and references therein for various ways of describing thermostats.

Instead of considering mechanical systems, it is simpler to model the interactions 
with the heat baths by a stochastic term in the equations of motion of the system 
(like in a Langevin equation). The microscopic dynamics becomes then stochastic.
This means that the evolution is given by a Markov chain with transition matrix $W(C',C)$
which represents the rate at which the system jumps from a configuration
$C$ to  a configuration $C'$ (i.e. the probability that the system jumps
from $C$ to $C'$ during an infinitesimal time interval $dt$ is given by
$W(C',C)dt$).

At equilibrium, one usually requires that the transition matrix satisfies {\it detailed balance}
\begin{equation}
W(C',C) \ e^{-{E(C) \over T}}
=W(C,C') \ e^{-{E(C') \over T}}
\label{detailed-balance}
\end{equation}
which ensures 
the time reversal symmetry of 
the microscopic dynamics. 
If one introduces $q$ the energy transfered from the heat bath at
temperature $T$ to the system,  and $W_q(C',C)dt$, the probability that
the system jumps during $dt$ from $C$ to $C'$ by receiving an energy $q$
from the heat bath, one can rewrite (\ref{detailed-balance})
\begin{equation}
e^{q \over T}
\ W_q(C',C) 
=W_{-q}(C,C') 
\label{detailed-balance-bis}
\end{equation}

If one accepts that detailed balance gives a good description of the coupling with a single  heat bath at temperature $T$, one can wonder what
would be the right way of describing the dynamics of a system coupled to several heat baths at unequal temperatures like in figure 1.
When the system jumps from one configuration $C$  to another configuration $C'$,  energies $q_a, q_b, q_c ...$ are transfered from the heat baths at temperatures $T_a, T_b,T_c ...$ to the system. The straightforward generalization of (\ref{detailed-balance-bis})  is 
\begin{equation}                               
e^{{q_a \over T_a}+ 
{q_b \over T_b}+
{q_c \over T_c}+
 ...}
\ W_{q_a,q_b,q_c..}(C',C)
=W_{-q_a,-q_b,-q_c...}(C,C')
\label{generalized-detailed-balance}                                                    \end{equation}
For a system in contact with several reservoirs at temperature $T_a,T_b,T_c...$,
this simply means,  by comparing with (\ref{detailed-balance-bis}), that
the exchange of energy with the heat bath at temperature
$T_a$ tend to equilibrate the system at temperature $T_a$,   the
exchange with the heat bath at temperature $T_b$ tend to equilibrate
the system at temperature $T_b$ and so on.

\medskip

The fluctuation theorem 
\cite{ECM,GC} can be easily recovered from 
the {\it generalized detailed balance relation} (\ref{generalized-detailed-balance}).
To see this, one can compare the probability of a
trajectory in phase space and its time reversal for a system in contact with two reservoirs.
Similar approaches have been implemented for stochastic dynamics in \cite{K,LS,M1,M2}.
A trajectory $"Traj"$ is specified by a sequence of successive
configurations $C_1,... C_k$ visited by the system,
the times $t_1,...t_k$ spent in each of these configurations, and the energies $
q_{a,i}, q_{b,i}$ transfered from the heat baths to the system  when the
system jumps from $C_i$ to $C_{i+1}$.
\[ {\rm Pro}(Traj) =  dt^{k-1}  \ 
\left[\prod_{i=1}^{k-1}  W_{q_{a,i}, q_{b,i}} (C_{i+1},C_i) \right]
 \ \exp \left[- \sum_{i=1}^{k} t_i  \ r(C_i)    \right] \]
where $r(C) = \sum_{C'} \sum_{q_a,q_b} W_{q_{a}, q_{b}} (C',C) $ and $dt$ is the infinitesimal
time interval over which jumps occur.

For the trajectory $"-Traj"$ obtained  from $"Traj"$ by time reversal,
i.e. for which the system visits successively the configurations
$C_k,... C_1$,  exchanging the energies $-q_{a,i}, -q_{b,i}$ each time
the system jumps from $C_{i+1}$ to $C_i$, one has
\[ {\rm Pro}(-Traj) =  dt^{k-1}  \
\left[\prod_{i=1}^{k-1}  W_{-q_{a,i}, -q_{b,i}} (C_{i},C_{i+1}) \right]
 \ \exp \left[- \sum_{i=1}^{k} t_i \ r(C_i)    \right] \]
One can see from the generalized detailed balance relation
(\ref{generalized-detailed-balance}) that
\begin{equation}
 {{\rm Pro}(Traj) \over {\rm  Pro}(-Traj) } = \exp \left[ - \sum_{i=1}^{k-1}
{q_{a,i} \over T_a} +  {q_{b,i} \over T_b} \right] = \exp \left[ -
{Q_t^{(a)} \over T_a} -  {Q_t^{(b)} \over T_b} \right]
\label{ratio}
\end{equation}
where $Q_t^{(a)}= \sum_i q_{a,i} $ and $Q_t^{(b)}= \sum_i q_{b,i} $
are the total energies transfered from the heat baths $a$ and $b$ to the
system during time $t$.
If the internal energy of the system is bounded, energy conservation implies that $|Q_t^{(a)} + Q_t^{(b)}| < E$,  and one gets
\begin{equation}
 \exp \left[ Q_t^{(a)} \left({1  \over T_b} -  {1 \over T_a} \right) - {E
\over T_b} \right] < 
 {{\rm Pro}(Traj) \over  {\rm Pro}(-Traj)}  < 
 \exp \left[ Q_t^{(a)} \left({1  \over T_b} -  {1 \over T_a} \right) + {E \over T_b} \right]
\label{ratio-bis}
\end{equation}

If $P(C)$ is the steady state probability of
configuration $C$, the probability that  $Q_t \equiv Q_t^{(a)}$ is the total energy transfered
from the heat bath $a$ to the system is given by
\[{\rm Pro}(Q_t) = \sum_{C_{\rm initial}} \sum_{C_{\rm final}}  \sum_{Traj(C_{\rm initial},C_{\rm final},Q_t)}  
 P(C_{\rm initial}) \  
{\rm Pro}( Traj(C_{\rm initial},C_{\rm final},Q_t)) \] where the sums are over all initial
configurations $C_{\rm initial}$, final configurations $C_{\rm final}$ and all trajectories
$Traj(C_{\rm initial},C_{\rm final},Q_t)$ starting in configuration $C_{\rm initial}$, ending in
configuration $C_{\rm final}$ with a total transfer of energy $Q_t$.
Now as
\[{\rm Pro}(-Q_t) = \sum_{C_{\rm initial}} \sum_{C_{\rm final}}  \sum_{Traj(C_{\rm final},C_{\rm initial},-Q_t)}  
 P(C_{\rm final}) \
{\rm Pro}( Traj(C_{\rm final},C_{\rm intial},-Q_t)) \]
one can see  that, if  for   any pair of configurations 
 the ratio of their steady state weights remains bounded
\[\forall C,C' \ \ \ 0 <  A < { P(C) \over P(C')} < B < \infty  \ , \]
one has because of  (\ref{ratio-bis}) that 
\[
 {A \over B} \exp \left[-{E \over T_b }+Q_t \left( {1 \over T_b} - {1 \over
T_a} \right)  \right]
< {{\rm Pro}(Q_t)  \over {\rm Pro}(-Q_t) } < 
 {B \over A} \exp \left[{E \over T_b }+Q_t \left( {1 \over T_b} - {1 \over T_a} \right) 
\right]
 \]
Taking the log and then the long time limit (\ref{F(j)}) leads to the {\it fluctuation
theorem}
\begin{equation}
\cF(j) - \cF(-j) = -j \left( {1 \over T_b} - {1 \over T_a}\right) 
\label{fluctuation-theorem-1}
\end{equation}
which states that the difference $\cF(j) - \cF(-j)$ is linear in $j$ with a universal slope
related to the difference of the inverse temperatures.

We see that in the framework of stochastic dynamics,
the fluctuation theorem is an elementary consequence of the generalized
detailed balance relation (\ref{generalized-detailed-balance}) satisfied
by the dynamics and of the assumptions that the energy is  bounded (see
 \cite{farago,HRS,visco} for examples where the energy is not bounded in which case the fluctuation theorem has to be modified) and the fact that the time $t$ is much longer than the relaxation times in the system.
In terms of the Legendre transform (\ref{generating-function},\ref{mu(lambda)})
the fluctuation theorem becomes
\begin{equation}
\mu(\lambda)=  \mu\left( - \lambda + {1 \over T_a} - {1 \over T_b}\right) 
\label{fluctuation-theorem-2}
\end{equation}
\ \\
{\it Remarks:}
\begin{enumerate}
\item 
In the limit of small $T_a- T_b$ (i.e. close to equilibrium), one can recover from 
(\ref{fluctuation-theorem-2}) the fluctuation-dissipation relation between the variance of the current at equilibrium 
\begin{equation}
 {\langle Q_t^2 \rangle \over t } \to \tilde \sigma \ \ \ \ \ \ {\rm for} \  T_a=T_b
\label{sigma-def}
\end{equation}
and
the response to a small temperature gradient
\begin{equation}
 {\langle Q_t \rangle \over t } \to (T_a - T_b) \tilde D \ \ \ \ \ \ {\rm for} \  T_a-T_b  \ {\rm small}
\label{D-def}
\end{equation}
In fact from these definitions of $\tilde \sigma$ and $\tilde D$, one has
\begin{equation}
 \mu(\lambda) =   (T_a- T_b) \tilde D \lambda  + {\tilde \sigma \over 2 } \lambda^2 + O \left( \lambda^3,  \lambda^2 (T_a - T_b),  \lambda (T_a - T_b)^2 \right)
\label{fluctuation-dissipation}
\end{equation}
and   for this expression to satisfy the fluctuation theorem (\ref{fluctuation-theorem-2}), the coefficients $\tilde \sigma$ and $\tilde D$ have to satisfy
\begin{equation}
\tilde \sigma = 2 T_a^2 \tilde D
\label{fluctuation-dissipation-bis}
\end{equation}
 which is the usual Einstein fluctuation-dissipation relation
between the response coefficient $\tilde D$ and the fluctuation coefficient
$\tilde \sigma$.
Note that in general both $\tilde D$ and $\tilde \gs$ depend on the temperature $T_a$.
\item One can easily extend the generalized detailed balance (\ref{generalized-detailed-balance}) and the fluctuation theorem (\ref{fluctuation-theorem-1},\ref{fluctuation-theorem-2}) to other types of currents. For example, in the case of a
current of particles, (\ref{generalized-detailed-balance}) becomes 
\begin{equation}
z_a^{-q_a} \; 
z_b^{-q_b }
\ W_{q_a,q_b}(C',C)
=W_{-q_a,-q_b}(C,C')
\label{generalized-detailed-balance-bis}
        \end{equation}
where $z_a$ and $z_b$ are the fugacities associated to the reservoirs of
particles and $q_a$ and $q_b$ are the numbers of particles transfered from the reservoirs while the system jumps from configuration $C$ to configuration $C'$.
The fluctuation theorem (\ref{fluctuation-theorem-1},\ref{fluctuation-theorem-2}) becomes then 
\begin{equation}
\label{fluctuation-theorem-3}
\cF(j) - \cF(-j) = j   [ \log z_b - \log  z_a ]
\qquad {\rm and} \qquad 
\mu(\lambda)=  \mu\left( - \lambda +  \log z_b  - \log z_a \right) 
\end{equation}
Close to equilibrium, if one defines as in (\ref{sigma-def},\ref{D-def}),  the fluctuation and the response coefficients for a system in contact with two reservoirs 
\begin{equation}                             
{\langle Q_t^2 \rangle \over t } \to \tilde \sigma \ \ \ \ \ \ {\rm for} \  \rho_a=\rho_b   
\label{sigma-def-bis}  
\qquad {\rm and} \qquad 
 {\langle Q_t \rangle \over t } \to (\rho_a - \rho_b) \tilde D \ \ \ \ \ \ {\rm for} \  \rho_a-\rho_b  \ {\rm small} 
\end{equation}
where $\tilde D$ and $\tilde \gs$ are now functions of the density $\gr_a$.
One can show, by expanding in powers of $\lambda$ and of $z_a-z_b$ as in (\ref{fluctuation-dissipation}) that
\begin{equation}  
 \tilde  \sigma = 2 \tilde D {d \rho \over d \log z}= 2 \tilde D T \rho^2 \kappa 
\label{fluctuation-dissipation-ter}
\end{equation}                       
where $\kappa = \rho^{-1} d \rho / dp$ is the compressibility.
(To see why the compressibility appears,  one can write $\log Z  =-\bbF/T = -V f(N/V)/T$  where $\bbF$ is the free energy and $f$ the free energy per unit volume; one uses the facts that $T \log z = d \bbF/dN= f'(\rho) $ and that
$p= - d\bbF/dV= \rho f'(\rho) - f(\rho)$; then one can see that $d \rho /d \log z=T/f"(\rho)=T \rho d\rho / dp$).
\item Another easy extension is to   consider systems with
 several types of currents (for example a current of particles and a current of energy, or several types of particles, or systems 
in contact with more than two reservoirs). 
The extension of the fluctuation theorem to these cases allows one to
recover Onsager's reciprocity  relations in the close-to-equilibrium limit \cite{G,M1}. 
\item
The fluctuation theorem is usually  formulated in terms of entropy production \cite{G2,ES,M2}
as, in the steady state, the entropy of the system remains stationary
whereas   a current $j$ of energy from the heat bath at
temperature $T_a$ into a heat bath at temperature $T_b$  gives a rate
 $j ( 1/T_b - 1/T_a)$ of increase of entropy. 
\end{enumerate}

\medskip

{\it An example: the symmetric simple exclusion process (SSEP)} \cite{Liggett,spohn,KL,dls}
\\ There are only few examples of non-equilibrium steady states for which
the cumulants $\langle Q_t^n \rangle_c $ or the large deviation function  $\cF(j)$  of the
current can be calculated \cite{ddr,hrs,wr}.
\begin{figure}[ht]
\centerline{{\includegraphics[width=7cm]{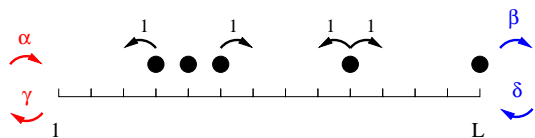}}}
\caption{The symmetric simple exclusion process }
\label{ssep}
\end{figure}
One of the simplest cases  is the symmetric simple exclusion process  shown in figure \ref{ssep}.

The model is defined as a one dimensional lattice of $L$ sites with open boundaries, each site being either
occupied by a single particle or empty.  During every infinitesimal time interval $dt$, each particle has a probability $dt$ of jumping to the left if the neighboring site on its left is empty, $dt$ of jumping to the right if the neighboring site on its right is empty. At the two boundaries the dynamics is modified to mimic the coupling with reservoirs of particles: at the left boundary, during each time interval $dt$, a particle is injected on site $1$ with probability $\alpha dt$ (if this site is empty) and a particle is removed from site $1$ with probability $\gamma dt$ (if this site is occupied). Similarly on site $L$, particles are injected at rate $\delta$ and  removed at  rate $\beta$.
(Note that one could consider that in the SSEP the particles represent
quanta of energy and all the properties could be interpreted in terms of
heat transport.)

From  the definition of the model, it is immediate to see that the dynamics satisfies the generalized detailed balance relation (\ref{generalized-detailed-balance-bis})
with \[ z_a = {\alpha \over \gamma} \ \ \ \ ; \ \ \ \ 
 z_b = {\delta \over \beta}  \]
If $\tau_i$ is a binary variable  which indicates whether site $i$ is occupied ($\tau_i=1$) or empty ($\tau_i=0$), it is easy to calculate the steady state profile \cite{ddr}
\begin{equation}
 \langle \tau_i \rangle = \rho_b + {L-i + b \over L
+ 1 + a + b} (\rho_a -
\rho_b)
\label{profile}
\end{equation}
 where 
\[ 
\rho_a = {\alpha \over \alpha + \gamma} 
\ \ \ \ , \ \ \ 
 \rho_b = {\delta \over \beta + \delta} 
\ \ \ \ \  {\rm and} \ \ \ \ 
a = {1 \over \alpha + \gamma} 
\ \ \ \ , \ \ \ 
 b = {1 \over \beta + \delta}  \ .
\]
Clearly, in the expression (\ref{profile}) of the profile, $\rho_a$ and $\rho_b$ represent the
densities  in the reservoirs at the two ends of the chain.
The calculation of the first cumulants can be done either directly or  by a perturbation theory in $\lambda$ by using the generating function \cite{ddr}.
\begin{eqnarray*}
 \lim_{t \to \infty} {\langle Q_t \rangle \over t}
 = {\rho_a - \rho_b
\over L + a +b -1} 
\end{eqnarray*}
\begin{eqnarray*}
 \lim_{t \to \infty}{   \langle Q_t^2 \rangle_c \over t} =
 {1
\over
L_1}( \rho_a + \rho_b - 2 \rho_a \rho_b)
 + {a (a-1) (2 a -1) + b
(b-1) (2b-1) - L_1 (L_1-1)(2 L_1-1) \over 3 L_1 ^3 (L_1-1)}
(\rho_a - \rho_b)^2
\end{eqnarray*}
where   $L_1=L+a+b-1$.
For large L, the first four cumulants are given by
\begin{eqnarray}
&&\lim_{t \to \infty}{   \langle Q_t \rangle \over t} \simeq {\rho_a -
\rho_b \over L} \label{cum1} \nonumber \\
&& \lim_{t \to \infty}{   \langle Q_t^2 \rangle_c \over t} \simeq {1 \over
L}\left[ \rho_a
+ \rho_b  -   {2 \rho_a^2 +2  \rho_a \rho_b + 2 \rho_b^2 \over 3} \right]
\nonumber \\
&&
\lim_{t \to \infty}{   \langle Q_t^3 \rangle_c \over t}  \simeq {1 \over
L}\left[ \rho_a - \rho_b  - 2(\rho_a^2 - \rho_b^2 ) 
 + {16\rho_a^3 +12 \rho_a ^2 \rho_b - 12 \rho_a \rho_b^2 -16 \rho_b^3
 \over 15 } 
\right] \nonumber \\
&& \lim_{t \to \infty}{   \langle Q_t^4 \rangle_c \over t} \simeq {1 \over
L} \left[ \rho_a + \rho_b  - {14 \rho_a^2  + 2 \rho_a \rho_b  + 14 \rho_b^2
 \over 3 } + 
  		 {32 \rho_a^3  +8 
\rho_a^2 \rho_b + 8 \rho_a \rho_b^2 + 32 \rho_b^3
\over 5}  \right. \nonumber \\  
&& \qquad \qquad \qquad \qquad \qquad \qquad \qquad \qquad \qquad \left.  
 - { 96 \rho_a^4  + 64 \rho_a^3 \rho_b  -40 \rho_a^2 \rho_b^2 + 64 \rho_a \rho_b^3
+96 \rho_b^4 \over 35} \right] 
\label{eq: cumulant 4}
\end{eqnarray}
\begin{figure}[ht]
\centerline{\includegraphics[width=9cm]{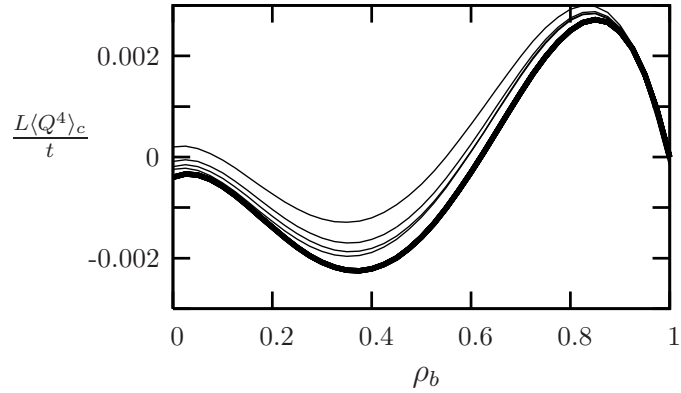}}
\caption{ The fourth cumulant versus $\rho_b$ for $\rho_a=1$.
The thin lines represent the fourth cumulant obtained from exact calculations of $\mu_L(\gl)$ for system sizes
$L=5,9,13,17$, whereas the thick line represents expression (\ref{eq: cumulant 4})
valid in the limit $L \to \infty$.}
\label{fig:4eme cumulant}
\end{figure}

\section{The additivity principle}
\label{sec: The additivity principle}

One can formulate a conjecture, {\it the additivity principle} \cite{bd}, based on a simple physical picture,
which allows one to determine all the cumulants and the large deviation function $\cF(j)$ for more general one dimensional diffusive systems.
Applied to the SSEP, this leads to the same expression of the cumulants  (\ref{eq: cumulant 4}) and provide 
a way of calculating all the higher cumulants.
Here we will limit the discussion to non-equilibrium steady states of systems in contact with two reservoirs of particles. As shown below everything can be easily generalized to systems in contact with two heat baths.
For a system of length $L+L'$  in contact with two
reservoirs of particles at densities $\rho_a$ and $\rho_b$, the
probability of observing, during a long time $t$,  an integrated current
$Q_t=jt$ has the following form (\ref{F(j)})
\begin{equation}
{\rm Pro}_{L+L'}\left(j,\rho_a,\rho_b\right) \sim e^{- t
F_{L+L'}\left(j,\rho_a,\rho_b\right)} \; .
\label{FL}
\end{equation}
The idea of the additivity principle is to relate the large deviation
function $F_{L+L'}(j,\rho_a,\rho_b)$ of the current to the large
deviation functions of subsystems of lengths $L$ and $L'$ by writing that for large $t$
\begin{equation}
{\rm Pro}_{L+L'}\left(j,\rho_a,\rho_b\right) \sim
\max_r \left[
{\rm Pro}_{L}\left(j,\rho_a, \ggr \right) \times
{\rm Pro}_{L'}\left(j, \ggr,\rho_b \right) \right] \; .
\label{PPP}
\end{equation}

This means that the probability of  transporting a current $j$ over a
distance $L+L'$ between two reservoirs at densities $\rho_a$ and $\rho_b$
is the same (up to boundary effects which give for large $L$ subleading
contributions) as the probability of transporting the same current $j$
over a distance $L$ between two reservoirs at densities
$\rho_a$ and $\ggr$ times the probability of transporting the current $j$
over a distance $L'$ between two reservoirs at densities $\ggr$ and
$\rho_b$. One can then argue that one should choose for $\ggr$ the
density which  makes this probability maximum.
From (\ref{PPP}) one gets the following additivity property of the large
deviation function
\begin{equation}
F_{L+L'}\left(j,\rho_a,\rho_b\right) =
\min_\ggr \left[
F_{L}\left(j,\rho_a, \ggr \right) + F_{L'}\left(j, \ggr ,\rho_b\right) \right] \; .
\label{FFF}
\end{equation}
By repeating this procedure, one gets that
\begin{equation}
F_L(j,\rho_a,\rho_b)= \min_{\ggr_1,... \ggr_{k-1}} \; \left \{\sum_{i=0}^{k-1} 
F_{l} (j, \ggr_i,\ggr_{i+1})  \right \}
\label{discrete}
\end{equation}
where $k=L/l, \ggr_0=\rho_a$ and $\ggr_k=\rho_b$.

\begin{figure}[ht]
\centerline{\includegraphics[width=5cm]{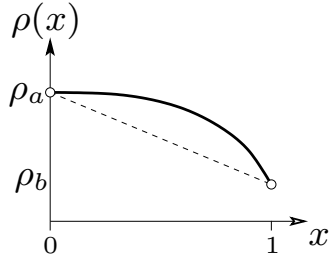}}
\caption{The dashed line represents the steady state profile.
The density changes in the bulk to facilitate the deviation of the current.}
\label{fig: decoupage}
\end{figure}

For  large  $L$ and $k$ (with $L/k$ still very large), if one considers current
fluctuations of order $1/L$, it is advantageous to minimize (\ref{discrete}) to make  
 the differences $\ggr_i - \ggr_{i+1}$  small.
As the current $j$ is also small 
one can consider that each piece of length $l$ is close to equilibrium and has Gaussian fluctuations at the 
leading order 
\begin{equation}
F_{l} (j, \ggr_i, \ggr_{i+1})
\simeq { [ j - {D(\ggr_i) (\ggr_i - \ggr_{i+1}) \over l}  ]^2 \over 2 {\sigma(\ggr_i) \over l}}  
\label{Fl}
\end{equation}
where the parameters $D$ and $\sigma$ are defined for a system of length $l$ in contact with two reservoirs at densities $\rho_a$ and $\rho_b$ by
\begin{equation}                             
{\langle Q_t^2 \rangle \over t } \to {\sigma(\rho_a) \over l} \ \ \ \ \ \ {\rm for} \  \rho_a=\rho_b   
\label{sigma-def-ter}  
\end{equation}        
\begin{equation}  
 {\langle Q_t \rangle \over t } \to (\rho_a - \rho_b) {D(\rho_a) \over l} \ \ \ \ \ \ {\rm for} \  \rho_a-\rho_b  \ {\rm small} 
 \label{D-def-ter}    
\end{equation}         
These are the same parameters as in (\ref{sigma-def},\ref{D-def},\ref{sigma-def-bis}) up to a factor $l$. 
(In the definitions (\ref{sigma-def-ter},\ref{D-def-ter}), one should take first the $t \to \infty$
limit, i.e.
$\sigma(\rho_a) = \lim_{l \to \infty} \lim_{t \to \infty} {l \langle Q_t^2 \rangle \over t }$
).

If for large $k$, the density $\ggr_i$ varies slowly with $i$ 
\[ \ggr_i = \rho \left(i { l\over L} \right) \]
for some smooth density $\gr(x)$ (see Figure \ref{fig: decoupage}) then combining (\ref{discrete}) and (\ref{Fl}), we get 
\begin{equation}
F_L(j,\rho_a,\rho_b)= 
\min_{ \{\ggr_i\} }  \  \sum_{i=0}^{k-1} 
{ [ j - {D(\ggr_i) (\ggr_i-\ggr_{i+1}) \over l}  ]^2 \over 2 {\sigma(\ggr_i) \over l}}
= 
\min_{\rho(x)}  \  
{1 \over L } \int_0^1  { [ Lj + \rho'(x) D(\rho(x))     ]^2 \over 2
\sigma(\rho(x)) } dx
\label{continuous}
\end{equation}
with $\rho(0)=\rho_a$ and $\rho(1)=\rho_b$. 
Note that (\ref{Fl}) is a local equilibrium assumption, i.e. that both the current $j$ and the difference $\ggr_i-\ggr_{i+1}$ are small. 
Therefore one cannot expect (\ref{continuous}) to be valid when the current deviation $j$ is not of order $1/L$.

The average profile $\overline{\rho}(x)$ is the one which makes vanish the large deviation function $F_L$. It therefore satisfies
$$L\overline{j}+D(\overline{\rho}(x)) \, \overline{\rho} (x)'=0 $$ 
where the most likely current $\overline{j}$ is fixed by the boundary conditions 
$$\overline{j}= {1 \over L} \int_{\rho_b}^{\rho_a} D(\rho) d \rho \, .$$

\section{The large deviation function obtained from the additivity principle}

\subsection{The optimal profile}

The profile $\rho_0(x)$ which optimizes (\ref{continuous}) satisfies 
\[ {d \over d\rho} {L^2 j^2 \over 2 \sigma(\rho_0(x))}
 -\rho_0'(x)^2 {d \over d\rho} {D^2(\rho_0(x))) \over 2 \sigma(\rho_0(x))}
 - 2 \rho_0''(x)    {D^2(\rho_0(x))) \over 2 \sigma(\rho_0(x))} =0 \]
If one multiplies this expression by $\rho_0'(x)$, one can   integrate once.  Finally one gets that the optimal profile satisfies
\begin{equation} 
\rho_0'(x)^2 ={(Lj)^2 \big( 1+2 K \sigma(\rho_0(x)) \big) \over D^2(\rho_0(x)) }
\label{profile-1s}
\end{equation}
where the  integration constant $K$  is fixed by the boundary conditions 
$\rho_0(0)=\rho_a$ and $\rho_0(1)=\rho_b$.

Suppose that $\gr_a>\gr_b$ and that the deviations are not too large so that 
the optimal profile remains monotone, i.e. 
\begin{equation}
\rho_0'(x)= - Lj { \sqrt{1+2 K \sigma(\rho_0(x))} \over D(\rho_0(x))}
\label{profile-bis}
\end{equation}
one can rewrite (\ref{continuous}) as
\begin{equation}
F_L(j,\rho_a,\rho_b)=  j \int_{\rho_b}^{\rho_a}  \left[ {1+ K \sigma(\rho) \over [1+2 K \sigma(\rho)]^{1/2}} -1 \right] {D(\rho) \over \sigma(\rho) }
\; d \rho
\label{LDJ}
\end{equation}
where the constant $K$ is fixed from (\ref{profile-bis}) by the boundary condition  ($\rho(0)=\rho_a$ and $\rho(1)=\rho_b$), i.e.
\begin{equation}                                                                
Lj= \int_{\rho_b}^{\rho_a}  { D(\rho) \over [1+2 K \sigma(\rho)]^{1/2}}   
 \; d \rho
  \label{LDJ_current}  
 \end{equation}             

The optimal profile (\ref{profile-bis}) remains unchanged when $j \to -j$ (simply  the sign of   $[1+2 K \sigma(\rho)]^{1/2}$ is changed)  in (\ref{LDJ},\ref{LDJ_current}) and one gets that
\begin{equation}
 F_L(j) - F_L(-j)= - 2 j \int_{\rho_b}^{\rho_a} {D(\rho) \over \sigma(\rho) } \; \d \rho
\label{GC}
\end{equation}
which is the fluctuation theorem (\ref{fluctuation-theorem-3}).
In fact 
already in (\ref{continuous}) it was clear by expanding the square that the
optimal $\rho_0(x)$ does not depend on the sign of $j$ and that (\ref{GC})
had to be satisfied.

\medskip

The physical meaning of the optimal profile $\gr_0$ (\ref{profile-1s}) is that adopting this profile is the easiest way to flow through the system an atypical current $j$.
The large deviation functional (\ref{continuous}) shows that the optimal density profile $\gr_0$ and the current deviation $j$ are coupled in a non trivial way. 
One can think of the system as a pipe with diameter $\gs(\gr)$ depending on the local density. 
The easiest way to increase the particle current is to adjust the size of the pipe $\gs(\gr)$
and therefore the local density, in order to facilitate the flow of particles.
In the example of the SSEP with reservoirs at equal densities $\gr = \gr_a =\gr_b$, 
the variance of the current $\gs(\gr) = 2 \gr (1-\gr)$  is maximum at density $\gr=1/2$.
When $\gr_a =\gr_b < 1/2$, it is favorable to have in the bulk a density $\gr_0(x) >\gr_a$ in order to facilitate
the flow of particles and the optimal way of doing it is by choosing the profile $\gr_0(x)$ which satisfies 
(\ref{profile-1s}).
If $\gr_a =\gr_b = 1/2$, then the optimal density profile remains flat for any current deviation and  the large deviation functional (\ref{continuous}) is quadratic \cite{ddr,bd}.
In general the complicated expression of the cumulants (\ref{eq: cumulant 4}) expresses the 
 non-trivial coupling between the flux $j$ and the optimal density profile $\gr_0(x)$.

\subsection{The cumulants}

The parametric expression (\ref{LDJ}, \ref{LDJ_current})
 for the large deviation function $F_L (j)$ can be transformed into another parametric form for $\mu_L(\lambda)$ defined in (\ref{generating-function},\ref{mu(lambda)})
\begin{equation}
\mu_L (\lambda,\rho_a,\rho_b) =
-   {K  \over L} \left[ \int_{\rho_b}^{\rho_a}  { D(\rho) \ d \rho
\over \sqrt{ 1 + 2 K \sigma(\rho)}} \right]^2  \, ,
\label{mu}
\end{equation}
with $K = K(\lambda,\rho_a,\rho_b)$ is the solution of
\begin{equation}
\lambda=
\int_{\rho_b}^{\rho_a} d \rho { D(\rho) \over
\sigma(\rho)}\left[{1
\over \sqrt{ 1 + 2 K \sigma(\rho)}} -1 \right] \, .
\label{lambda}
\end{equation}
By eliminating $K$, one can obtain the expansion of $\mu_L$ in powers of $\lambda$
 and by taking successive derivatives (\ref{mu(lambda)},\ref{cum}) with respect to $\lambda$  one gets for the cumulants of the current:
\begin{eqnarray}
\label{eq: cumulants}
&& {\langle Q_t \rangle \over t} = {1 \over L} I_1,
\quad
 {\langle Q_t^2 \rangle - \langle Q_t
\rangle^2
\over t} = {1 \over L} {I_2\over I_1}, \\ 
&& {\langle Q_t^3 \rangle_{ c} \over t}=  {1 \over L} { 3 (I_3 I_1 -
I_2^2) \over I_1^3} , \quad
{\langle Q_t^4 \rangle_{ c} \over t}=  {1 \over L} { 3 ( 5 I_4 I_1^2  - 14
I_1 I_2 I_3 +  9 I_2^3  ) \over I_1^5} \,  \nonumber
\end{eqnarray}
where the integrals $I_n$ are given by 
$$I_n = \int_{\rho_b}^{\rho_a} D(\rho)\  \sigma(\rho)^{n-1} \  d \rho \, .$$

\medskip

In the case of the SSEP, one has $ D(\rho)=1$ and $\sigma(\rho)=2 \rho(1-\rho)$.
One can simplify (\ref{mu},\ref{lambda}) and get
\begin{eqnarray}
\label{eq: mu SSEP}
\mu_L (\lambda)=  {1 \over L}\left[\log(\sqrt{1+ \omega} + \sqrt{\omega} ) \right]^2 
\quad {\rm with} \quad \omega = (e^\lambda -1) \rho_a + (e^{-\lambda}-1) \rho_b - (e^\lambda -1)  (e^{-\lambda}-1)  \rho_a \rho_b. 
\end{eqnarray}
From this one can recover the cumulants (\ref{eq: cumulant 4}) and determine all the higher cumulants.
Expressions equivalent to (\ref{eq: mu SSEP}) were derived in the theory of shot noise of mesoscopic conductors \cite{BB,LLY}.

\medskip

\noindent
{\it Remark.} 
The density $\gr$ is the physical relevant parameter. However it can be useful \cite{HS} to consider instead the conjugate field $\gb = \log z$ (see (\ref{fluctuation-dissipation-ter})). Formula (\ref{continuous}) then
simplifies as it depends only on one macroscopic input $\gs(\gb)$ 
\begin{equation}
F_L(j,\gb_a,\gb_b)=  
\min_{\gb(x)}  \  
{1 \over L } \int_0^1  { [ Lj + {\gs(\gb(x)) \over 2} \gb'(x) ]^2 \over 2
\sigma(\gb(x)) } dx
\label{continuous conjugate}
\end{equation}
where the minimum is taken over the chemical potential profiles such that 
$\gb(0)=\log z_a$ and $\gb(1)=\log z_b$.

\subsection{Heat flux.}

All the above discussion can be generalized to the case of a heat flux in diffusive systems: one has to replace everywhere the density profile $\rho(x)$ by the temperature profile $T(x)$. There is even one simplification as in the thermal case 
$D$ and $\sigma$ are related as in (\ref{fluctuation-dissipation-bis}) so
that for $D(T)$ defined as in (\ref{D-def-ter}) (i.e. $\langle Q_t \rangle/t= (T_a-T_b) D(T)/L$ for small $T_a- T_b$), one gets
\begin{eqnarray}
Lj= \int_{T_b}^{T_a} {d T  D(T) \over [1+ 4 K T^2 D(T)]^{1/2}} \, , 
\qquad \qquad 
F_L (j)=j  \int_{T_b}^{T_a} \, {dT \over 2 T^2} \, \left[ {1+ 2 K T^2  D(T) \over [1+ 4 K T^2 D(T)]^{1/2}} - 1 \right]
\label{eq: heat}
\end{eqnarray}

\section{The macroscopic fluctuation theory}
\label{sec: macroscopic fluctuation}

Building on the hydrodynamic large deviation theory \cite{KOV,spohn,KL}, Bertini et al 
developed \cite{BDGJL1,BDGJL2,BDGJL3} a general framework to determine 
 the steady state large deviation function of non equilibrium systems. 
This framework has been extended \cite{bdgjl,BDGJL4} to the current large deviations.
Let us sketch briefly their approach. 
For diffusive systems (such as SSEP), the total flux $Q_i(t)$ flowing through position $i$ between 
time $0$ and time $t$ and the density $\rho_i(t)$ near position $i$ are, for a large system of size $L$
and for times of order $L^2$, scaling functions of the form
\[ Q_i(t) = L \hat Q \left( {i\over L}, {t \over L^2} \right), 
\qquad {\rm and} \qquad  
 \rho_i(t) =  \hat \rho \left( {i\over L}, {t \over L^2} \right) \,  .\]
It is convenient to introduce the instantaneous current defined in terms of the rescaled time $\tau$
\begin{equation}
\hat q(x,\tau) = {\partial \hat Q (x,\tau) \over \partial \tau} 
\end{equation}
In fact $L \hat q(x,\tau) d \tau$ is simply the total flux of particles through position $[x L]$ during the microscopic time interval
$[L^2 \tau, L^2 (\tau+d\tau)]$, with $1/L \ll d\tau \ll 1$ so that there is a large number of particles which contribute to the integrated current but the density does not vary over this small time interval.
Remark that the current $\hat q$ is defined after a diffusive rescaling, i.e. the space is scaled by $1/L$ and the time by $1/L^2$. Thus unlike the microscopic current, $\hat q$ remains of order $1$.
The conservation of the number of particles implies that
\begin{equation} {
\partial \hat \rho(x, \tau) \over \partial \tau}=
-{\partial^2 \hat Q (x,\tau) \over \partial \tau \partial x }
= -{\partial \hat q(x,\tau) \over   \partial x } 
\label{conservation}
\end{equation}
The {\it macroscopic fluctuation theory} \cite{bdgjl,BDGJL4} gives for the probability of observing
a certain density profile $\hat \rho \left( x, \tau \right)$ and a current $\hat q \left( x, \tau \right)$
over the rescaled time interval $0 < \tau' < \tau$
\begin{equation}
\label{eq: dev exp}  
{\rm Pro} \Big( \{\hat \rho(x,\tau'), \hat q(x,\tau')\}  \Big)  \sim \exp \left[ - L \int_0^\tau  
d \tau' \int_0^1 dx {\left[\hat q(x,\tau') + D({\hat \rho(x,\tau'}))
{\partial {\hat \rho(x,\tau')} \over \partial x}\right]^2 \over 2 \sigma(\hat \rho(x,\tau'))} \right] 
\end{equation}
Of course $\hat \rho$ and $\hat q$ have to satisfy the relation (\ref{conservation}).
(Note that if $t$ is the microscopic time, then $\tau = t/ L^2$ plays the role of a macroscopic time).
A similar expression was obtained in \cite{pjsb,jsp} 
by considering stochastic models in the context of shot noise in mesoscopic quantum conductors.
The functional (\ref{eq: dev exp})  was used to calculate the large deviation functional of the density
for several systems \cite{BDGJL2,BGL} and in the case of SSEP the results agree with an exact microscopic
derivation \cite{dls2}.

\medskip

The large deviation function $\cF(j)$ (\ref{F(j)}) for observing the total current $j$, i.e. the following event
\begin{equation}
L j = {1 \over \tau}  \int_0^{\tau} d \tau' \int_0^1 dx\; \hat q(x,\tau')
\label{constraint}
\end{equation}
as predicted  in \cite{bdgjl,BDGJL4} by the macroscopic fluctuation theory (\ref{eq: dev exp}) becomes  
\begin{equation}
\cF(j) =  {1 \over  L} \  \lim_{\tau \to \infty} \ {1 \over \tau} \min_{\hat \rho(x,\tau') \atop \hat q(x,\tau')}
\int_0^{\tau} d \tau' \int_0^1 dx {\left[\hat q(x,\tau') +
D({\hat \rho(x,\tau'})) {\partial {\hat \rho(x,\tau')} \over \partial x}\right]^2 \over 2 \sigma(\hat \rho(x,\tau'))}
\label{Bertini}
\end{equation}
 where  the minimun is over all the density profiles
$\{\hat \rho(x,\tau'), 0 < \tau' < \tau \}$ and the current
$\{\hat q(x,\tau'), 0 < \tau' < \tau \}$ which satisfy the conservation law (\ref{conservation}) and the global constraint
(\ref{constraint}).

If the optimal density and current profiles are time independent (up to boundary effects for $\tau'$ close to 0 or 
$\tau$ which do not contribute in the $\tau \to \infty$ limit), one
recovers the predictions of the additivity principle (\ref{continuous}) and $\cF(j) = F_L(j)$.
When the optimal profile is time dependent  the  additivity principle predictions  (\ref{continuous}) give only 
an  upper bound : $\cF(j) \leq F_L (j)$.
In \cite{bdgjl,BDGJL4}, Bertini et al provided an example for which the functions $F_L$ and $\cF$
are different. In their example, $F_L$ was not a convex function of $j$ and $\cF$ was its
convex envelope (see  (\ref{convexity})).
They also proved (see \cite{BDGJL4} Section 6.1) that $\cF$ reduces to $F_L$ under the  following 
global condition on $D$ and $\gs$
\begin{eqnarray}
\label{eq: no transition}
{\rm For \ all} \ \gr, \qquad \qquad \qquad D(\gr) \gs''(\gr) \leq D'(\gr) \gs'(\gr)
\end{eqnarray}
This holds for the SSEP and the Zero Range process \cite{KL}.
Condition (\ref{eq: no transition}) is however only  sufficient. 

In general  a dynamical phase transition \cite{BDGJL4,bd2} may occur where the system switches  
from a time independent to a time dependent optimal profile.
To calculate the large deviation function $\cF$ one
needs to determine the optimal time dependent profile
$\hat \rho(x,\tau)$, which is not an easy task as the optimization problem is non-linear.
A  complete characterization  of the regime for which the additivity principle holds 
($\cF(j) =F_L(j)$) remains a challenging problem.

\section{Phase transitions}
\label{sec: Phase transition}

In this section, we try to determine the phase boundary where the optimal profile becomes time dependent.
To do so we consider a more general situation with a small driving force (like an electric field)
in the bulk.
Adding such a driving force of amplitude $\nu/L$ to an open system of length $L$ with reservoirs $\gr_a,\gr_b$ modifies the mean current (\ref{D-def-ter}) as follows \cite{spohn,bd2}
\begin{equation}  
 {\langle Q_t \rangle \over t } \to (\rho_a - \rho_b) {D(\rho_a) \over L} 
+{\nu \over L} \gs(\gr_a) \qquad
\qquad {\rm for} \  \rho_a-\rho_b\ {\rm small} 
\label{eq: current weak}
\end{equation}     
In (\ref{eq: current weak}), the conductivity $\gs$ which was defined as the variance of the current in (\ref{sigma-def-ter})
can also be understood as the linear response to the small field $\nu/L$.
The effect of the field can be easily taken into account in the framework of the macroscopic fluctuation theory \cite{BDGJL4,bd2}
by arguing that locally the current has Gaussian fluctuations with mean value given by (\ref{eq: current weak}).
The functional (\ref{Bertini}) becomes
\begin{equation}
\cF(j) = {1 \over L} \ 
\lim_{\tau \to \infty} \ {1 \over \tau} \min_{\hat \rho(x,\tau') \atop \hat q(x,\tau')}
\int_0^{\tau} d \tau' \int_0^1 dx {\left[\hat q(x,\tau') + 
D({\hat \rho(x,\tau'})) {\partial {\hat \rho(x,\tau')} \over \partial x}
- \nu \gs({\hat \rho(x,\tau'})) \right]^2 \over 2 \sigma(\hat \rho(x,\tau'))} \, .
\label{LD}
\end{equation}
with the constraint (\ref{constraint}) on the total current.

	\subsection{Stability of the functional}

The time independent optimal profile $\rho_0(x)$ is now a solution of (see (\ref{profile-1s}))
\begin{equation}
\big( D(\gr_0(x)) \gr_0^\prime (x) \big)^2
= \big( jL - \nu \gs (\gr_0(x)) \big)^2 + 2 K \gs (\gr_0(x)) \, ,
\label{eq: optimal nu}
\end{equation}
where the constant $K$ has to be adjusted so that $\gr_0$ satisfies the boundary conditions
($\gr_0(0)=\gr_a$ and $\gr_0(1)=\gr_b$).
A situation  for which $\rho_0(x)$ is certainly not
optimal in (\ref{LD}) is when a small time dependent perturbation is sufficient to lower (\ref{LD}). 
To investigate the stability of $\rho_0(x)$ against such perturbations, one can write
\[ \hat \rho (x,\tau')= \rho_0(x) + \delta \rho(x,\tau')  \]
\[ \hat q (x,\tau')= j + \delta j(x,\tau')  \]
where  $\delta \rho$ and $\delta j$  have zero time averages and are related by (\ref{conservation}).
Inserting these expressions into (\ref{LD}), one gets at the second order
\begin{eqnarray*}
\int_0^{\tau} d \tau' \int_0^1 dx  \left\{
{(\delta j)^2 \over 2 \sigma(\gr_0)}  - j {\sigma'(\gr_0) \over \sigma^2(\gr_0)} \delta \rho \, \delta j 
+ A(\gr_0) (\delta \rho')^2 
+ 2 \Big[ A'(\gr_0) \gr_0^\prime \Big] \; \delta \rho \, \delta \rho'
+ \frac{1}{2} \Big[ B''(\gr_0) + A''(\gr_0) (\gr_0^\prime)^2 \Big] (\delta  \rho)^2 
 \right\} \nonumber 
\end{eqnarray*}
where we introduced the functions
\begin{eqnarray*}
A(u) ={D^2(u) \over 2 \sigma(u)}, 
\quad {\rm and} \quad 
B(u) = {j^2 \over 2 \sigma(u)}  + {\nu^2 \over 2} \sigma(u) \; . 
\end{eqnarray*}

The coefficients of the quadratic form are $x$ dependent but time independent.
Thus to analyze the stability, one can consider perturbations of the form
\begin{eqnarray} 
&& \delta \rho(x,\tau') = {i \over  \go} \exp(i \go \tau') \gp_1 '(x) - {i \over  \go}   \exp(- i \go \tau') \gp_2 '(x)
\nonumber \\
&& \delta j(x,\tau')  = \exp(i \go \tau') \gp_1 (x) +    \exp(- i \go \tau') \gp_2 (x)
\label{eq: perturbation}
\end{eqnarray}
where $\gp_2(x) =\gp_1^* (x)$.
The quadratic form can be rewritten as 
\begin{eqnarray}
\label{2nd var}
&& \tau \;  \int_0^1 dx  \left\{
{\gp_1 \, \gp_2 \over  \sigma(\gr_0)}  - 
j {\sigma'(\gr_0) \over \sigma^2(\gr_0)} 
{i \over \go} \left[ \gp_1' \, \gp_2 -\gp_1 \, \gp_2' \right]
+ {2  \over \go^2} A(\gr_0)  \gp_1 '' \, \gp_2 '' \right.\\
&& \qquad \qquad  \left. 
+ {2 \over \go^2} \;  \Big[ A'(\gr_0) \gr_0^\prime \Big] \,  \left[ \gp_1' \, \gp_2'' +\gp_1'' \, \gp_2' \right]
+ \frac{1}{\go^2} \Big[ B''(\gr_0) + A''(\gr_0) (\gr_0^\prime)^2 \Big] \gp_1' \, \gp_2' 
 \right\} \nonumber 
\end{eqnarray}
For the time independent profile $\gr_0(x)$ to be stable against small time dependent perturbations, the quadratic
form (\ref{2nd var}) has to be positive for all $\go$. 
In general, all the coefficients in (\ref{2nd var}) are spatially dependent through $\gr_0(x)$ and it is difficult to provide 
from (\ref{2nd var}) a more explicit  characterization of this local stability.

	\subsection{Periodic systems}

For a system of $N$ particles on a ring of length $L$ with density $\bar \gr=N/L$, 
the flat profile $\gr_0(x) = \bar \gr$ remains a solution of the Euler-Lagrange equation associated 
to (\ref{LD}). 
In this case, the coefficients of the quadratic form have no $x$ dependence and 
 the different spatial modes decouple. One can choose 
$\gp_1 = \exp( i \, kx)$ and $\gp_2 = \exp( -i \, kx)$ and
the positivity of the  quadratic form implies that for any  $k$ (multiple of $2 \pi$) and $\go$, one has
\begin{eqnarray}
{1 \over \gs(\bar \gr)} \left(1 + Lj {k  \gs'(\bar \gr) \over \go \gs(\bar \gr)} \right)^2
 + {k^2  \over \go^2} \left({k^2  D(\bar \gr)^2 \over \gs(\bar \gr)} 
-(Lj)^2 {  \gs''(\bar \gr) \over 2 \gs(\bar \gr)^2} + \nu^2 {\sigma''(\bar \gr) \over 2 }\right) > 0.
\label{eq: stability wave}
\end{eqnarray}
The first mode to become unstable is the fundamental mode $k = 2 \pi$, thus
the flat profile is stable when \cite{bd2}  
\begin{eqnarray}
 {8 \pi^2 D^2(\bar \gr) \over \sigma(\bar \gr)} 
>  \gs''(\bar \gr) \Big[  {(Lj)^2 \over \sigma^2(\bar \gr)} - \nu^2  \Big] \, .
\label{eq: critique1}
\end{eqnarray}
Let $j_c$ be the critical current for which (\ref{eq: critique1}) becomes an equality.
If $\gs''(\bar \gr)>0$ (resp $\gs''(\bar \gr)<0$), the flat profile becomes unstable for currents $|j| > |j_c|$
(resp $|j| < |j_c|$).
Remark that the instability regime is always symmetric with respect to 0 as predicted by
the fluctuation theorem (\ref{fluctuation-theorem-3}) which becomes in presence of a driving force
\begin{eqnarray*}
\cF(j) - \cF(-j) = - 2 \nu j - 2 j \int_{\gr_b}^{\gr_a} {D(\gr) \over \gs(\gr)} \, d \gr  \; .
\end{eqnarray*}
Beyond the threshold (\ref{eq: critique1}) a bifurcation occurs and a traveling wave of the form 
$\gr(x- v t)$ is more favorable than the flat profile $\bar \gr$. 
From (\ref{eq: stability wave}), we get that close to the phase transition, the optimal velocity
is given by 
\begin{eqnarray}
\go = - 2 \pi L j_c {\gs'(\bar \gr) \over \gs(\bar \gr)}
\qquad \Rightarrow \qquad
v =  L j_c {\gs'(\bar \gr) \over \gs(\bar \gr)}
\label{eq: velocity}
\end{eqnarray}

\bigskip

If we make the {\it assumption} that no first order transition occurs before the second order transition 
predicted at $j_c$, we can compute the expansion of $\cF$ close to  $j_c$.
We consider small current perturbations
\begin{eqnarray}
Lj = L j_c + \eps,
\qquad {\rm with} \qquad
L j_c =  \sqrt{\nu^2 \sigma(\bar \gr)^2 + {8 \pi^2 D(\bar \gr)^2 \sigma(\bar \gr) \over \gs''(\bar \gr)}}
\label{eq: j critique}
\end{eqnarray}
Let us limit the discussion to the case $\gs''(\bar \gr) <0$. Then for $\gep>0$ the flat profile 
remains optimal and one expects that the large deviation function is quadratic 
\begin{eqnarray}
\label{eq: gaussien}
\forall \gep>0, \qquad  \cF(j) =
{\left(L j_c + \gep -\nu {\gs(\bar \gr)}   \right)^2 \over 2 \gs(\bar \gr)}
\end{eqnarray}
On the other hand for $\gs''(\bar \gr) <0$, the flat profile is unstable for $\gep<0$ and
the expansion of $\cF(j)$ at the second order in $\gep$ can be obtained by approximating
the travelling wave as follows
\begin{eqnarray}
\label{eq: approx wave}
\gr_0 (x,t) \approx \bar \gr + \sqrt{-\gep} \, a_1 \sin \big( 2 \pi (x - v t) \big)
+ \gep \, a_2 \cos \big(4 \pi (x - v t) \big)
\end{eqnarray}
where the velocity $v = L j_c {\gs'(\bar \gr) \over \sigma(\bar \gr)}$ is given by (\ref{eq: velocity}).
In fact the expansion should also include corrections in $\gep$ to the velocity as well as other Fourier modes, but a computation shows that they do not contribute to the second order expansion of $\cF$.
Inserting the test function (\ref{eq: approx wave}) in  (\ref{LD}) implies that the second order of $\cF(j)$
in $\gep$  is given by the quartic form
\begin{eqnarray} 
&& \left(
\frac{Lj_c {\gs''(\bar \gr)}}{4 {\gs (\bar \gr)}^2}
+ \left( - \frac{3 \pi ^2 \,{D'(\bar \gr)} {D(\bar \gr)}}{ {\gs (\bar \gr)}}
+  \frac{ 3 \pi ^2 \,{D(\bar \gr)}^2 {\gs'(\bar \gr)} }{2 {\gs (\bar \gr)}^2 }
+ \frac{ \pi ^2 \, {D(\bar \gr)}^2 {\gs^{(3)} (\bar \gr)} }{2 {\gs (\bar \gr)} {\gs''(\bar \gr)}} \right) a_2 \right) a_1^2 
\nonumber \\ 
&& + \left(-\frac{\pi^2  {D(\bar \gr)}^2 \,  {\gs^{(4)} (\bar \gr)}}{16 \, {\gs (\bar \gr)} {\gs''(\bar \gr)}} 
+\frac{\pi ^2  {D(\bar \gr)}^2 {\gs'(\bar \gr)}^2}{4 {\gs (\bar \gr)}^3}
+\frac{\pi ^2 {D(\bar \gr)}^2 \, {\gs''(\bar \gr)} }{4 {\gs (\bar \gr)}^2}
-\frac{ \pi ^2 {D(\bar \gr)} {D'(\bar \gr)} \, {\gs'(\bar \gr)} }{2 {\gs (\bar \gr)}^2} \right. \nonumber\\
&& \qquad \qquad \qquad \qquad \left.
+\frac{\pi^2 {D(\bar \gr)} {D''(\bar \gr)} }{4 {\gs (\bar \gr)}}+\frac{\pi^2 {D'(\bar \gr)}^2 }{4 {\gs (\bar \gr)}}
+\frac{3 \nu^2 {\gs''(\bar \gr)}^2}{64 {\gs (\bar \gr)}}  \right) a_1^4 
+ \frac{3 \pi^2 {D(\bar \gr)}^2  }{{\gs (\bar \gr)}} a_2^2+\frac{1}{2 {\gs (\bar \gr)}}   
\label{eq: quartic}
\end{eqnarray}
where $\gs^{(3)},\gs^{(4)}$ denote the third and fourth derivatives.
The optimal amplitudes $a_1,a_2$ of the traveling wave (\ref{eq: approx wave}) are the minimizers of 
the quartic form. Note that (\ref{eq: quartic}) is not always stable and for some specific choices
of the functions $D$ and $\gs$, the mimimum of (\ref{eq: quartic}) can be $- \infty$. For example the sign of the coefficient 
$a_1^4$ depends on ${\gs^{(4)} (\bar \gr)}$ which can a priori take  any arbitrary value. 
The condition for the quartic form (\ref{eq: quartic}) to be stable can be written as
\begin{eqnarray}
\label{eq: quartic >0}
\bbA(\bar \gr) >0 \, ,
\end{eqnarray}
where 
\begin{eqnarray}
&& \bbA(\bar \gr) =
9 \nu ^2 {\gs(\bar \gr)}^2 {\gs''(\bar \gr)}^4
- 96 \pi^2 {D'(\bar \gr)}^2 {\gs(\bar \gr)}^2 {\gs''(\bar \gr)}^2 
+ 48 \pi^2  {D(\bar \gr)} {D''(\bar \gr)} {\gs(\bar \gr)}^2 {\gs''(\bar \gr)}^2  \nonumber  \\
&& \qquad \qquad 
+ 12 \pi^2 {D(\bar \gr)}^2 {\gs'(\bar \gr)}^2 {\gs''(\bar \gr)}^2
- 24 \pi^2 {D(\bar \gr)}^2 {\gs(\bar \gr)} {\gs'(\bar \gr)} {\gs''(\bar \gr)}  {\gs^{(3)}(\bar \gr)} \nonumber  \\
&& \qquad \qquad
+ 48 \pi^2 {D(\bar \gr)}^2 {\gs(\bar \gr)} {\gs''(\bar \gr)}^3
+ 48 \pi^2 {D(\bar \gr)} {D'(\bar \gr)} {\gs(\bar \gr)}  {\gs'(\bar \gr)} {\gs''(\bar \gr)}^2 
+ 48 \pi^2 {D(\bar \gr)} {D'(\bar \gr)} {\gs(\bar \gr)}^2 {\gs''(\bar \gr)} {\gs^{(3)}(\bar \gr)} \nonumber \\
&& \qquad  \qquad 
- 4 \pi^2 {D(\bar \gr)}^2 {\gs(\bar \gr)}^2  {\gs^{(3)}(\bar \gr)}^2  - 12 \pi^2 {D(\bar \gr)}^2 {\gs(\bar \gr)}^2   {\gs''(\bar \gr)} {\gs^{(4)}(\bar \gr)} \, . 
\label{eq: A}
\end{eqnarray}
If (\ref{eq: quartic >0}) is not satisfied then one expects that a first order transition occured before $j_c$.

\medskip

We suppose as before that $\gs''(\bar \gr)<0$ and that (\ref{eq: quartic >0}) is satisfied. Then the minimum of
(\ref{eq: quartic}) is achieved for
\begin{eqnarray}
a_1 =
2  \sqrt{ - \frac{6 L j_c \,  {\gs(\bar \gr)} {\gs''(\bar \gr)}^3  }{ \bbA(\bar \gr)}}
\qquad {\rm and} \qquad
a_2 = \left( {{D'(\bar \gr)} \over 2 {D(\bar \gr)}} -    {{\gs'(\bar \gr)} \over 4 {\gs (\bar \gr)}}-    {{\gs^{(3)} (\bar \gr)} \over 12 {\gs''(\bar \gr)}} \right) a_1^2
\end{eqnarray}
where $\bbA(\bar \gr)$ is defined in (\ref{eq: A}).
Thus for $\gep<0$, one finds at the second order
\begin{eqnarray}
\label{eq: non gaussien}
\cF(j) =
{\left(L j_c + \gep  -\nu {\gs(\bar \gr)}   \right)^2 \over 2 \gs(\bar \gr)}
-   \frac{3 {\gs''(\bar \gr)}^4 (L j_c)^2}{\gs(\bar \gr) \, \bbA(\bar \gr)}  \gep^2
+ O \big( \gep^3 \big) \, .
\end{eqnarray}
Comparing (\ref{eq: gaussien}) to (\ref{eq: non gaussien}), we see that if $\bbA(\bar \gr)>0$
the time dependent profile gives lower $\cF (j)$ than the flat profile.

\medskip

As an example, we consider the weakly asymmetric simple exclusion process (WASEP) which follows the same exclusion rule as the SSEP with jump rates biased by $ \exp(\nu/L)$ to the right and $ \exp(-\nu/L)$ to the left.
In this case $D(\gr)=1$ and $\gs(\gr)= 2 \gr (1-\gr)$. 
Since $\gs''(\bar \gr) = -4$, the threshold of stability of the flat profile is given by 
\begin{eqnarray*}
L j_c =  \sqrt{\nu^2 \sigma(\bar \gr)^2 - {2 \pi^2 \sigma(\bar \gr) }}
\end{eqnarray*}
For $\gep<0$, a perturbation of the form (\ref{eq: approx wave}) leads to the quartic form
(\ref{eq: quartic})
\begin{eqnarray}
\label{eq: quartic WASEP}
 {\pi^2 + {3 \over 4} \big( \nu^2 \gs(\bar \gr)^2 - 4  \pi^2 \gs(\bar \gr) \big) \over \gs(\bar \gr)^3} \, a_1^4
+ { 3  \pi^2 \gs' (\bar \gr) \over  2 \gs(\bar \gr)^2}a_1^2 \, a_2 
+ {  3  \pi^2  \over \gs(\bar \gr)}  a_2^2
- {L j_c \over \gs(\bar \gr)^2} a_1^2 + {1 \over 2 \gs(\bar \gr)}
\end{eqnarray}
The quartic form (\ref{eq: quartic WASEP}) is stable so that the amplitudes of 
the traveling wave (\ref{eq: approx wave}) are given by 
\begin{eqnarray}
a_1 = \sqrt{ {2 L j_c \gs(\bar \gr) \over 3 \nu^2 \gs(\bar \gr)^2 - 6 \pi^2\gs(\bar \gr)  + \pi^2}}
\qquad {\rm and} \qquad
a_2 = - { \gs'(\bar \gr) \over 4 \gs(\bar \gr)} a_1^2
\end{eqnarray}
Finally for $\gep<0$, one finds at the second order
\begin{eqnarray*}
\cF(j) =
{\left(L j_c + \gep  - \nu {\gs(\bar \gr)}   \right)^2 \over 2 \gs(\bar \gr)}
-  { \nu^2  \gs(\bar \gr) - 2 \pi^2 \over 3 \nu^2 \gs(\bar \gr)^2 - 6 \pi^2\gs(\bar \gr)  + \pi^2}  \gep^2 
+ O \big( \gep^3 \big) \, .
\end{eqnarray*}

It is interesting to note that for the WASEP on a ring at density $\bar \gr = 1/2$, 
the optimal velocity is 0, thus the optimal
profile remains time independent. 
This means that the phase transition could have been detected already at the level
of the functional $F_L$.

\bigskip

Finally, let us mention that in the large drift limit $\nu \to \infty$, the asymptotic cost for (\ref{LD}) as well as the asymptotic
shape of the optimal traveling waves can be  computed \cite{bd2}.
In particular for the WASEP, the 
current large deviation function (\ref{LD}) converges in the large 
drift limit $\nu \to \infty$ to the current 
large deviation function of the totally asymmetric simple exclusion process \cite{dl,da}.
We refer the reader to \cite{bd2} for further details and to \cite{bd3} for a study of the large drift limit
 in the case of open systems.


\section{Conclusion}

In this paper, we have shown how to generalize the detailed balance relation to take into account the effect 
of reservoirs. From this generalized detailed balance relation the fluctuation theorem \cite{ECM,GC}
can be recovered which characterizes the odd part of the large deviation function of the current.
By a simple additivity principle \cite{bd}, one can predict for diffusive systems the whole large deviation function as well as all the 
cumulants of the current.
These predictions  agree with previous exact computations for some stochastic models like the symmetric simple exclusion process \cite{ddr}.
For some models, however, the additivity principle provides only an upper bound of the large deviation function of the current
\cite{bdgjl}.
This fact as well as the occurrence of phase transitions has been discussed in the framework of the macroscopic fluctuation
theory \cite{BDGJL4,bd2}.

A challenging issue would be to characterize precisely the range of validity of the additivity principle in the case of diffusive stochastic models.
Here we were only able to address the local stability of the time independent solution of the additivity principle.
How to calculate the large deviation function of the heat or particle current in a more general framework
(several species of particles, additional conserved or non conserved quantities) is also an interesting open
question.




\begin{thebibliography}{00}






\bibitem{LLP} S. Lepri, R. Livi, A. Politi,
Thermal conduction in classical low-dimensional lattices, {\it  Phys. Rep.} {\bf    377},  1-80 (2003)


\bibitem{ruelle1} D. Ruelle,
Conversations on nonequilibrium physics with an extraterrestrial, {\it Physics Today} {\bf 57}, No 5, 48-53 (2004)

\bibitem{ruelle2} D. Ruelle,
Smooth dynamics and new theoretical ideas in nonequilibrium statistical mechanics, {\it J. Statist. Phys.}
{\bf 95}, 393-468 (1999)



\bibitem{DF}
A. De Masi, P. Ferrari,
A remark on the hydrodynamics of the Zero-Range Processes, {\it J. Stat. Phys.},
{\bf 36}, 81--87 (1984)


\bibitem{KLS}
S. Katz, J. Lebowitz, H.Spohn,
Non-equilibrium Steady States of Stochastic Lattice Gas Models of Fast Ionic Conductors,
{\it J. Stat. Phys.} {\bf 34}, 497-537 (1984)


\bibitem{DEHP}
B. Derrida, M.R. Evans, V. Hakim, V. Pasquier, 
Exact solution of a 1d asymmetric exclusion model using a matrix formulation,
{\it J. Phys. A} {\bf 26}, 1493-1517 (1993)

\bibitem{SD}
G. Sch{\"u}tz, E. Domany,
Phase transitions in an exactly soluble one-dimensional asymmetric exclusion
 model, {\it J. Stat. Phys.} {\bf 72}, 277-296 (1993)

\bibitem{DZ} 
A. Dembo, O. Zeitouni, Large deviations techniques and applications. Second edition. {\it Applications of Mathematics}, {\bf 38}. Springer-Verlag, (1998)

\bibitem{E} 
R. Ellis, Entropy, large deviations, and statistical mechanics, {\it Reprint of the 1985 original. Classics in Mathematics Springer-Verlag}, Berlin, (2006)


\bibitem{DSt1}
M. Depken, R. Stinchcombe,
Exact Joint Density-Current Probability Function for the Asymmetric 
Exclusion Process,
{\it Phys. Rev. Lett.} {\bf 93}, 040602 (2004)

\bibitem{DSt2}
M. Depken, R. Stinchcombe,
Exact probability function for bulk density and current in the asymmetric exclusion process,
{\it Phys. Rev. E} {\bf 71}, 036120 (2005)

\bibitem{bdgjl} L.  Bertini, A.  De Sole, D.  Gabrielli, G.
Jona--Lasinio, C.  Landim, Current Fluctuations in Stochastic Lattice Gases,
{\it Phys.  Rev.  Lett.}~{\bf 94}, 030601 (2005)

\bibitem{dls3}
 B. Derrida, J.L. Lebowitz, E.R. Speer, Exact free energy functional for a driven diffusive open stationary nonequilibrium system  {\it Phys. Rev. Lett.} {\bf 89}, 030601 (2002)

\bibitem{EPR} 
J.-P. Eckmann, C.-A. Pillet, L. Rey-Bellet, 
Non-equilibrium statistical mechanics of anharmonic chains coupled to two heat baths at different temperatures, {\it Comm. Math. Phys.}  {\bf 201},  no. 3, 657--697  (1999)

\bibitem{BLR} F. Bonetto, J.L. Lebowitz, L. Rey-Bellet,
Fourier's law: a challenge to theorists, Mathematical Physics 2000,
128-150,  {\it Imperial College Press} (2000).  math-ph/0002052


\bibitem{ECM} 
D.J. Evans, E.G.D. Cohen, G.P. Morriss, 
Probability of second law violations  in shearing steady states,
{\it Phys. Rev. Letts.} {\bf 71}, 2401 (1993)

\bibitem{GC}    
G. Gallavotti, E.D.G. Cohen, Dynamical ensembles in stationary states, {\it J. Stat. Phys.} {\bf 80}, 931-970 (1995)

\bibitem{K}
J. Kurchan, Fluctuation Theorem for stochastic dynamics,  {\it J.  Phys.} {\bf A31} 3719,  (1998)



\bibitem{LS}   
J.L. Lebowitz, H.~Spohn, A Gallavotti-Cohen Type Symmetry
in the Large Deviation Functional for Stochastic Dynamics
{\it J.   Stat. Phys.} {\bf 95}, 333-366 (1999)

\bibitem{M1}
C. Maes, The fluctuation theorem as a Gibbs property, {\it J. Stat. Phys.} {\bf 95}, 367-392  (1999)

\bibitem{M2}
C. Maes, On the origin and the use of fluctuation relations for the entropy, {\it S\'eminaire Poincar\'e} 
{\bf 2},  29-62 (2003).

\bibitem{farago}
J.  Farago,
Power fluctuations in stochastic models of dissipative systems, 
{\it Physica A} {\bf 331},  69-89  (2004)

\bibitem{HRS}
R. J. Harris, A. Rakos, G. M. Sch\"utz,
Breakdown of Gallavotti-Cohen symmetry for stochastic dynamics,
{\it Europhysics Letters}, {\bf 75}, 227 - 233  (2006)

\bibitem{visco}
P. Visco, 
Work fluctuations for a Brownian particle between two thermostats, 
{\it J. Stat.  Mech.}  P06006 (2006)




\bibitem{G}    
G. Gallavotti, Chaotic hypothesis: Onsager reciprocity and fluctuation--dissipation theorem,  {\it J. Stat. Phys.} {\bf 84},  
899--926, (1996)

\bibitem{G2}    
G. Gallavotti,
Entropy production in nonequilibrium thermodynamics: a point of view, {\it Chaos} {\bf 14}, 680--690, (2004)

\bibitem{ES} 
D.J. Evans, D.J. Searles, The Fluctuation Theorem, {\it Advances in Physics} {\bf 51}, 1529-1585 (2002)


\bibitem{Liggett}   
T. Liggett, 
{\em Stochastic interacting systems: contact, voter and exclusion processes},
Fundamental Principles of Mathematical Sciences, {\bf 324} Springer-Verlag, Berlin, (1999)

\bibitem{spohn} H. Spohn,
{\em Large scale dynamics of interacting particles}, Springer (1991)

\bibitem{KL} C. Kipnis, C. Landim,
{\em Scaling limits of interacting particle systems}, Springer (1999)

\bibitem{dls} B.  Derrida, J.  L.  Lebowitz, E.  R.  Speer, Free energy
functional for nonequilibrium systems: an exactly solvable case,
{\it Phys. Rev. Lett.}  {\bf 87}, 150601 (2001)

\bibitem{ddr}
B. Derrida, B. Dou\c{c}ot, P.-E. Roche,
Current fluctuations in the one-dimensional symmetric exclusion process
with open boundaries, {\it J.  Stat. Phys.}  {\bf 115}, 717-748 (2004)


\bibitem{hrs}
 R. J. Harris, A. R\'akos, G. M. Sch\"utz,
Current fluctuations in the zero-range process with open boundaries,
{\it J. Stat. Mech.}  P08003 (2005)


\bibitem{wr} F. van Wijland, Z. R\'acz,  Large deviations in weakly
interacting boundary driven lattice gases,
{\it   J.   Stat. Phys.} {\bf  118}, 27-54  (2005)

\bibitem{bd}
T.  Bodineau, B. Derrida,
 Current fluctuations in nonequilibrium diffusive systems: An additivity
principle {\it Phys. Rev. Lett.} {\bf 92},  180601 (2004)

\bibitem{BB}
Y.M. Blanter, M. B\"uttiker,  Shot noise in mesoscopic conductors, {\it Phys. Rep.} {\bf 336},
1-166  (2000)
 
\bibitem{LLY}
Hyunwoo Lee, L. S. Levitov, A. Yu. Yakovets, Universal statistics of
transport in disordered conductors, {\it Phys. Rev. B} {\bf  51}, 4079-4083 (1995)

\bibitem{HS} H. Spohn, {\it Private communication}


\bibitem{KOV} C. Kipnis, S. Olla, S. Varadhan, Hydrodynamics and
large deviations for simple exclusion processes, {\it Commun. Pure
Appl. Math.} {\bf 42}, 115-137 (1989)


\bibitem{BDGJL1} L.  Bertini, A.  De Sole, D.  Gabrielli, G.
Jona--Lasinio, C.  Landim, Fluctuations in stationary non equilibrium states of
irreversible processes, {\it Phys.  Rev.  Lett.}~{\bf 87}, 040601 (2001)

 \bibitem{BDGJL2} L.  Bertini, A.  De Sole, D.  Gabrielli, G.
Jona--Lasinio,
C.  Landim, Macroscopic fluctuation theory for stationary non equilibrium
states,  {\it J. Stat, Phys.} {\bf 107},  635-675  (2002)
 

\bibitem{BDGJL3} L.  Bertini, A.  De Sole, D.  Gabrielli, G.
Jona--Lasinio,
C.  Landim,
Large deviations for the boundary driven symmetric simple exclusion
process,
 Math. Phys. Analysis  and  Geometry {\bf 6},  231-267  (2003)


\bibitem{BDGJL4}
L.  Bertini, A. De Sole, D. Gabrielli, G. Jona-Lasinio, C. Landim,
Non equilibrium current fluctuations in stochastic lattice gases,  
{\it J. Stat. Phys.}  {\bf 123},  no. 2, 237--276  (2006)


\bibitem{BGL}   L.  Bertini,  D.  Gabrielli, J. Lebowitz,
Large deviation for a stochastic model of heat flow, {\it J. Stat. Phys.} {\bf 121}, no. 5-6, 843-885 (2005)


\bibitem{dls2} B.  Derrida, J.  L.  Lebowitz, E.  R.  Speer, Large
deviation of the density profile in the symmetric simple exclusion
process,
{\it J. Stat. Phys.} {\bf 107}, 599--634 (2002)



\bibitem{pjsb}
S. Pilgram, A.N. Jordan, E.V. Sukhorukov, M.  Buttiker,
Stochastic path integral formulation of full counting statistics,
{\it Phys. Rev. Lett.}  {\bf 90},  206801  (2003)

\bibitem{jsp}
A.N. Jordan, E.V. Sukhorukov, S. Pilgram,
 Fluctuation statistics in networks: A stochastic path integral approach,
{\it  J. Math. Phys.} {\bf  45} 4386-4417  (2004)

\bibitem{bd2}
T.  Bodineau, B. Derrida,
Distribution of current in nonequilibrium diffusive systems and phase transitions,
{\it Phys. Rev. E} (3)  {\bf 72},  no. 6, 066110   (2005)


\bibitem{dl} B. Derrida, J.L.  Lebowitz,
Exact large deviation function in the asymmetric exclusion process,  {\it Phys. Rev. Lett.}  {\bf 80}, 209-213 (1998)

\bibitem{da}
B. Derrida, C. Appert, Universal large deviation function of the
Kardar-Parisi-Zhang equation in one dimension {\it J. Stat. Phys.} {\bf 94}, 1-30 (1999)

\bibitem{bd3}
T.  Bodineau, B. Derrida, 
Current large deviations for asymmetric exclusion processes with open boundaries,
{\it  J. Stat. Phys.} {\bf 123},  no. 2, 277--300  (2006)



\end{thebibliography}
\end{document}